# Quantum Mechanical Approach for Modeling of Ternary Based Strained-Layer Superlattice

Arash Dehzangi[1,*], Jiakai Li[2]

[1]Department of Electrical and Computer Engineering, University of Texas at Dallas, Richardson, Texas, USA.
[2] Applied Optoelectronics Inc, Sugar Land, Texas, 77478, USA
*arash.dehzangi@utdallas.edu

*Abstract*— **Ternary-based InAs/InAs$_{1-x}$Sb$_x$ Strained-Layer Superlattice (SLS) material with type-II band alignment belongs to the 6.1 Å family with reasonably small lattice mismatch with GaSb substrate for epitaxial growth. InAs/InAs$_{1-x}$Sb$_x$ SLS have been proven to have more advantages such as longer carrier lifetime, better control on growth and manufacturability, and being considered as an alternative material system for infrared photodetectors. In this article a quantum mechanical based modelling on electronic band structure of InAs/InAs$_{1-x}$Sb$_x$ is presented. A modified sp3s* empirical tight binding method along with implementing a virtual crystal approximation with a bowing of the s-on-site tight-binding energy, were incorporated. In this approach, a theoretical explanation of atomic segregation in superlattices is suggested and used in calculations. The simulations show good agreement with experimentally measured band gap of InAs/InAs$_{1-x}$Sb$_x$ superlattices.**

Strained-Layer Superlattice based on type-II superlattices band-structure [1-4] have attracted a lot of attention during last decade as an advanced metamaterial for several applications such as photodetectors [5, 6], laser[7], space application [8] and transistors [9-12]. The spatial separation of the electrons and holes into different quantum wells allows for great flexibility in bandgap engineering [13]. At first promising results reported for InAs/GaSb SLS [14-17] for different devices with the great flexibility for infrared detection and imaging [18-20] from short-wavelength infrared [21-24] (SWIR) to very long wavelength infrared [25] (VLWIR) regime. but eventually InAs/InAs$_{1-x}$Sb$_x$ SLS have been proven to have more advantages such as longer carrier life-time [26-30], better control on growth and manufacturability, and being considered as an alternative material system for infrared photodetectors. Gallium has been proven to be a major source of defects in InAs/GaSb SLS degrading the minority carrier lifetime.

While both InAs/InAs$_{1-x}$Sb$_x$ and InAs/GaSb SLS material sharing the same type-II band alignment, they hold certain differences in superlattice design challenges. InAs and GaSb both belong to the 6.1 Å family with small lattice mismatch, which makes it easier for a device designer to come up with a superlattice design. On the contrary, InAs and InAs$_{1-x}$Sb$_x$ are not closely lattice-matched, limiting the freedom to design. Any change in constituent layer thickness or/and the Sb composition (x) in the InAs$_{1-x}$Sb$_x$ layer must be carefully considered in selection of layer thicknesses. These limitations can affect the electronic band structure modeling process of this kind of superlattices especially when antimony-rich (x > 0.5) InAs$_{1-x}$Sb$_x$ layers are used in the superlattice structure.

Some aspects of the InAs/InAs$_{1-x}$Sb$_x$ SLS have been modeled previously using k.p method [31-33]. Ab-initio calculation of the electronic structures is a powerful tool, yet issues like practicality and time-consuming process impeded the popularity of the approach. The empirical tight binding model (ETBM) has shown promising capability for accurate calculation of SLS material [34-38]. However, ETBM is not as straightforward for InAs/InAs$_{1-x}$Sb$_x$ SLS, as it is for the case of InAs/GaSb binary SLS [39, 40] due to complications arises from handling ternary compounds additional to binary ones. In InAs$_{1-x}$Sb$_x$ ternary compounds, the group V elements are randomly distributed inside the crystal making it difficult to locate the As and Sb atoms, which in turn makes it hard for calculation/fitting of the orbital interaction energies between each atom and adjacent neighbors. These energies are the core parameters for ETBM approach. The fitting parameters are highly dependent on material quality, which demands a universal rule to consistently explain the experimental results for superlattices grown under different conditions.

In this work, a quantum mechanical based modelling on electronic band structure modeling of ternary-based SLS is presented. A virtual crystal approximation (VCA) was used to model the ternary compounds and generate fitting parameters for the empirical tight-binding method. In addition, we considered a crucial parameter during the growth as the antimony segregation in the InAs layers. With the addition of this aspect, our model can be used to explain our experimental results for superlattices grown at different growth conditions, especially the bandgaps, within some growth uncertainties.

Slater and Koster [41] in 1954 proposed the tight-binding method based on Bloch's linear combination of atomic orbital (LCAO) to interpolate the results of first-principle electronic structure calculations. While the method was not mature enough for an empirical approach due to limitation in computational capability. The limitation of the model mostly arises from the need for evaluation of the large numbers of integrals due to the non-orthogonality of the atomic orbitals required for the interaction energies. Löwdin attempt to address the issue by proposing an orthonormalization[42] of the atomic orbitals to preserve the useful symmetries of the corresponding orbitals under certain conditions. These symmetries are imperative to build the Hamiltonian matrix for superlattices. Following this strategy, we consider a basis of sp3s* orbitals with spin-orbit interaction similarly to the approach was proposed by Wei et al [39] for InAs/GaSb superlattices. Since Vogl [43] proposed s* orbital, the optimization of sp3s* tight binding parameters has been extensively studied and provided more accuracy for the calculation of band structure with tight binding method. The foundation of theoretical approach in using ETBM for both bulk III-V materials and III-V superlattices was studied extensively by Harrison and Vogl [43, 44]. Here we start briefly on description of the formalism and then focus on the current work.

Assume that $\phi_n^\alpha$ is a sp3s* Löwdin orbital of the atom in the crystal, where n runs through all the atoms in a unit cell and $\alpha$ runs through all Löwdin orbitals (s, px, py, pz, and s*). Then, using the Bloch's wavefunction of the electrons inside the material as a combination of all the Löwdin orbitals:

$$\psi_{\vec{k}}(\vec{r}) = \sum_{\overrightarrow{R_{cell}}} \sum_{n=1}^{N} \sum_{\alpha} C_n^\alpha \exp\left(i\vec{k} \cdot (\overrightarrow{R_{cell}} + \overrightarrow{\tau_n})\right) \phi_n^\alpha(\vec{r} - \overrightarrow{R_{cell}} - \overrightarrow{\tau_n}) \quad (1)$$

Where k is the wave vector, $\overrightarrow{R_{cell}}$ is a vector labeling for the position of one unit cell, $\tau_n$ is the coordinate position vector of the atoms in the cell, and $C_n^\alpha$ are constants. The energy dispersion relation of the crystal can be reduced to the following eigenvalue problem ($H$ is the Hamiltonian):

$$H\psi_{\vec{k}}(\vec{r}) = E\psi_{\vec{k}}(\vec{r}). \quad (2)$$

Considering the orthonormality and the symmetry properties of the Löwdin orbitals, we can simplify the matrix form of (1). We also should note that the superlattice's Hamiltonian only considers the interaction energies between the nearest neighbor's orbitals.

Following similar approach for superlattice Hamiltonian proposed for InAs/GaSb SLS [39], we assumed that each layer of the superlattice begins with an anion and ends with a cation. The Hamiltonian for superlattice is written:

$$H_{superlattice} = \begin{bmatrix} H_a & H_{ac} & 0 & 0 & \cdots & 0 & H_{ca}^+ \\ H_{ac}^+ & H_c & H_{ca} & 0 & \cdots & 0 & 0 \\ 0 & H_{ca}^+ & H_a & H_{ac} & 0 & \cdots & 0 \\ 0 & 0 & H_{ac}^+ & H_c & 0 & \cdots & 0 \\ \vdots & \vdots & 0 & 0 & \ddots & 0 & 0 \\ 0 & 0 & \vdots & \vdots & 0 & H_a & H_{ac} \\ H_{ca} & 0 & 0 & 0 & 0 & H_{ac}^+ & H_c \end{bmatrix} \quad (3)$$

where

$$H_v = \begin{bmatrix} h_{vv} & h_{v,so} \\ h_{v,so}^+ & h_{vv}^* \end{bmatrix}, \quad v = a, c \quad (4)$$

$$h_{vv} = \begin{pmatrix} E_{sv} & 0 & 0 & 0 & 0 \\ 0 & E_{pv} & -i\frac{\Delta_v}{3} & 0 & 0 \\ 0 & i\frac{\Delta_v}{3} & E_{pv} & 0 & 0 \\ 0 & 0 & 0 & E_{pv} & 0 \\ 0 & 0 & 0 & 0 & E_{ssv} \end{pmatrix}, \quad 6)$$

$$h_{v,so} = \begin{pmatrix} 0 & 0 & 0 & 0 & 0 \\ 0 & 0 & 0 & \frac{\Delta_v}{3} & 0 \\ 0 & 0 & 0 & -i\frac{\Delta_v}{3} & 0 \\ 0 & -i\frac{\Delta_v}{3} & i\frac{\Delta_v}{3} & 0 & 0 \\ 0 & 0 & 0 & 0 & 0 \end{pmatrix}, \quad 7)$$

and

$$H_{ac} = \begin{bmatrix} h_{ac} & 0 \\ 0 & h_{ac} \end{bmatrix}, H_{ca} = \begin{bmatrix} h_{ca} & 0 \\ 0 & h_{ca} \end{bmatrix}, \quad (8)$$

$$h_{ac} = \beta \begin{pmatrix} f_1 E_{sasc} & \sqrt{3}f_2 \cos\theta_x V_{sasc} & \sqrt{3}f_2 \cos\theta_y V_{sasc} & \sqrt{3}f_1 \cos\theta_z V_{sasc} & 0 \\ -\sqrt{3}f_2 \cos\theta_x V_{sasc} & f_1[E_{sasc} + (3\cos^2\theta_x - 1)V_{sayc}] & 3f_1 \cos\theta_x \cos\theta_y V_{sayc} & 3f_2 \cos\theta_x \cos\theta_z V_{sayc} & -\sqrt{3}f_2 \cos\theta_x V_{sasc} \\ -\sqrt{3}f_2 \cos\theta_y V_{sasc} & 3f_1 \cos\theta_x \cos\theta_y V_{sayc} & f_1[E_{sasc} + (3\cos^2\theta_y - 1)V_{sayc}] & 3f_2 \cos\theta_y \cos\theta_z V_{sayc} & -\sqrt{3}f_2 \cos\theta_y V_{sasc} \\ -\sqrt{3}f_1 \cos\theta_z V_{sasc} & 3f_2 \cos\theta_x \cos\theta_z V_{sayc} & 3f_2 \cos\theta_y \cos\theta_z V_{sayc} & f_1[E_{sasc} + (3\cos^2\theta_z - 1)V_{sayc}] & -\sqrt{3}f_1 \cos\theta_z V_{sasc} \\ 0 & \sqrt{3}f_2 \cos\theta_x V_{ssasc} & \sqrt{3}f_2 \cos\theta_y V_{ssasc} & \sqrt{3}f_1 \cos\theta_z V_{ssasc} & 0 \end{pmatrix} \quad (9)$$

$$h_{ca} = \beta \begin{pmatrix} g_1 E_{sasc} & -\sqrt{3}g_2 \cos\theta_x V_{sasc} & \sqrt{3}g_2 \cos\theta_y V_{sasc} & \sqrt{3}g_1 \cos\theta_z V_{sasc} & 0 \\ \sqrt{3}g_2 \cos\theta_x V_{sasc} & g_1[E_{sasc} + (3\cos^2\theta_x - 1)V_{sayc}] & -3g_1 \cos\theta_x \cos\theta_y V_{sayc} & -3g_2 \cos\theta_x \cos\theta_z V_{sayc} & \sqrt{3}g_2 \cos\theta_x V_{ssasc} \\ -\sqrt{3}g_2 \cos\theta_y V_{sasc} & -3g_1 \cos\theta_x \cos\theta_y V_{sayc} & g_1[E_{sasc} + (3\cos^2 y - 1)V_{sayc}] & 3g_2 \cos\theta_y \cos\theta_z V_{sayc} & -\sqrt{3}g_2 \cos\theta_y V_{ssasc} \\ -\sqrt{3}g_2 \cos\theta_z V_{sasc} & -3g_2 \cos\theta_x \cos\theta_z V_{sayc} & 3g_2 \cos\theta_y \cos\theta_z V_{sayc} & g_1[E_{sasc} + (3\cos^2\theta_z - 1)V_{sayc}] & -\sqrt{3}g_1 \cos\theta_z V_{ssasc} \\ 0 & -\sqrt{3}g_2 \cos\theta_x V_{ssasc} & \sqrt{3}g_2 \cos\theta_y V_{ssasc} & \sqrt{3}g_1 \cos\theta_z V_{ssasc} & 0 \end{pmatrix} \quad (10)$$

Where

$$\beta = \frac{3}{(1+\varepsilon_{xx})^2 + (1+\varepsilon_{yy})^2 + (1+\varepsilon_{zz})^2}, \quad (11)$$

$$\cos\theta_j = \sqrt{\frac{\beta}{3}}(1+\varepsilon_{jj}), j = x, y, z. \quad (12)$$

$\varepsilon_{jj}$ are the strain tensor diagonal components of the superlattice. In our approach, we consider the strain stems from average lattice constant of the grown superlattice, instead of what it is generally assumed, which is extracting it from the lattice mismatch between the substrate and the grown layer. This approach was strongly supported by the high-resolution X-ray diffraction (HR-XRD) performed on the actual grown SLS samples [45, 46], where the diffraction pattern of a superlattice grown on a substrate shows at least two strong peaks, one for the substrate and one corresponding to the average lattice constant of the superlattice. Given to this assumption, the strain tensor coefficients can be written as

$$\varepsilon_{xx} = \varepsilon_{yy} = \frac{a_{AV}}{a_i} - 1 \quad (13)$$

where

$$a_{AV} = \frac{\sum_{i=1}^{N} t_i a_i}{\sum_{i=1}^{N} t_i} \quad (14)$$

is the average lattice constant of the superlattice [47]. N is the total number of different layers in the superlattice design, $t_i$ is the thickness of the i-th layer, and $a_i$ is the lattice constant corresponding to the i-th layer. Finally, the strain tensor component in the growth direction (001) are calculated as

$$\varepsilon_{zz} = -D_i^{001}\varepsilon_{xx} = -\frac{c_{12}}{c_{11}}\varepsilon_{xx} \quad (15)$$

where $c_{12}$ and $c_{11}$ are elastic constants of the i-th layer.

The remaining constants in the Hamiltonian superlattice are linked to the crystalline structure constituted by the successive layers of the superlattice:

$$f_1 = \frac{1}{4}[\exp(i\vec{k}\cdot\overrightarrow{\tau_1}) + \exp(i\vec{k}\cdot\overrightarrow{\tau_2})], \quad (16)$$
$$f_2 = \frac{1}{4}[\exp(i\vec{k}\cdot\overrightarrow{\tau_1}) - \exp(i\vec{k}\cdot\overrightarrow{\tau_2})],$$

$$g_1 = \frac{1}{4}[\exp(-i\vec{k}\cdot\vec{\tau_3}) + \exp(-i\vec{k}\cdot\vec{\tau_4})],$$
$$g_2 = \frac{1}{4}[\exp(-i\vec{k}\cdot\vec{\tau_3}) - \exp(-i\vec{k}\cdot\vec{\tau_4})].$$

where $\vec{\tau_i}$ are the vectors linking two adjacent atoms in the crystalline structure, one from group III cation and one from group V anion. $\vec{\tau_i}$ can be written under the effect of strain as

$$\vec{\tau_1} = \frac{1}{4}a_i \begin{pmatrix} 1+\varepsilon_{xx} \\ 1+\varepsilon_{yy} \\ 1+\varepsilon_{zz} \end{pmatrix}, \vec{\tau_2} = \frac{1}{4}a_i \begin{pmatrix} -1-\varepsilon_{xx} \\ -1-\varepsilon_{yy} \\ 1+\varepsilon_{zz} \end{pmatrix}, \quad (17)$$

$$\vec{\tau_3} = \frac{1}{4}a_i \begin{pmatrix} 1+\varepsilon_{xx} \\ -1-\varepsilon_{yy} \\ -1-\varepsilon_{zz} \end{pmatrix}, \vec{\tau_4} = \frac{1}{4}a_i \begin{pmatrix} -1-\varepsilon_{xx} \\ 1+\varepsilon_{yy} \\ -1-\varepsilon_{zz} \end{pmatrix}$$

where $a_i$ is the lattice constant of the binary III-V component constituted by two adjacent atoms. For ternary compounds, the $a_i$ lattice constant is simply replaced with the value obtained by Vegard law[48].

The empirical tight-binding parameters (TBPs) used in this work for binary components are based on Vogl's and Klimeck's parameters [43, 49]. These TBPs are mostly fitted for room temperature (300 K), for the present work Harrison's d-2 scaling rule [50-52] was implemented to adjust TBPs for modelling of the electronic band structure at 77 K. At lower temperature the mean bond length between atoms decreases leading to higher interaction energy between the orbitals of these atoms which are covalently bonded. For that purpose, the tight-binding parameters have been fitted by multiplication of the inter-atomic orbital interaction energies [40], to reflect the impact of temperature on the mean distance between a given anion/cation couple. Although the distance between an anion and cation atom in a compound semiconductor is not directly linked to the distance between the orbitals of these atoms, but Varshni's empirical law [53] enables our approach to model the effect of temperature on the simulated band gap around the Γ-point. The set of modified tight-binding parameters at 77K for the relevant binary compound semiconductors are given in Table I.

$V_{sasc}$, $V_{saxc}$, $V_{xasc}$, $V_{s*axc}$, $V_{xas*c}$, $V_{xaxc}$, $V_{xayc}$ parameters in Table I are interaction energies between the s, p and s* orbitals of anions/cations. They are absolute energies which can be directly used in the superlattice Hamiltonian. While, $E_{sa}$, $E_{sc}$, $E_{pa}$, $E_{pc}$, $E_{s*a}$, $E_{s*c}$ are relative energies and are fitted to set the top of the valence band at a reference energy origin. For accurate modeling the band alignment between different materials, it is crucial to find valence band offset (VBO) in addition to these parameters. ab-initio calculations was previously used finding VBO of electronic structure by individually model different semiconductor hetero-junctions [54]. Anderson [55] in much simpler approach for a semiconductor hetero-junction calculated the conduction band offset by the difference in the electron affinities of the two materials in the hetero-junction.

**Table I.** The TBPs of relevant binary compound semiconductors at 77 K in eV.

|  | InAs | InSb | GaAs | GaSb | AlAs | AlSb |
|---|---|---|---|---|---|---|
| $E_{sa}$ | -9.3562 | -9.3378 | -9.2664 | -6.0493 | -7.6201 | -4.5572 |
| $E_{sc}$ | -3.9611 | -3.3248 | -4.3504 | -4.0712 | -1.1786 | -4.1180 |
| $E_{pa}$ | +1.8201 | +0.39352 | +1.4866 | +0.91157 | +0.8905 | +0.01635 |
| $E_{pc}$ | +3.1842 | +2.0791 | +3.2136 | +2.6352 | +3.4939 | +4.8741 |
| $E_{s*a}$ | +7.0432 | +6.6378 | +8.7826 | +7.8753 | +7.3905 | +9.8429 |
| $E_{s*c}$ | +6.1232 | +5.3807 | +5.8765 | +4.8565 | +6.6339 | +7.4324 |
| $V_{sasc}$ | -6.5328 | -5.8320 | -7.9480 | -5.7762 | -6.7375 | -6.7066 |
| $V_{saxc}$ | +4.3563 | +4.1129 | +2.7777 | +4.4761 | +5.1668 | +4.6377 |
| $V_{xasc}$ | +7.0778 | +7.5769 | +10.001 | +8.2748 | +6.3693 | +8.6279 |
| $V_{s*axc}$ | +2.9977 | +3.4448 | +3.6271 | +5.0079 | +4.5713 | +7.4657 |
| $V_{xas*c}$ | +5.3966 | +5.8873 | +7.0071 | +6.3813 | +7.3803 | +6.3653 |
| $V_{xaxc}$ | +2.5466 | +1.2596 | +2.3069 | +1.8244 | +1.8987 | +1.1192 |
| $V_{xayc}$ | +5.4646 | +4.0026 | +5.0305 | +5.3733 | +3.9025 | +4.9535 |
| $\Delta_a$ | +0.420 | +0.973 | +0.420 | +0.973 | +0.350 | +0.704 |
| $\Delta_c$ | +0.393 | +0.393 | +0.174 | +0.174 | +0.024 | +0.0306 |

Other purely theoretical methods suggest that the band offset can be obtained by using calculations of bulk band structure on an absolute energy scale. We use the valence band offset suggested by Van de Walle and Martin[56] (Table II) since it presents average values compared to the other methods available and has been widely used for calculations of band offsets for different combinations of materials.

**Table II.** The VBO of relevant III-V compound semiconductor which have been used in the superlattice Hamiltonian.[56]

|  | InAs | InSb | GaAs | GaSb | AlAs | AlSb |
|---|---|---|---|---|---|---|
| VBO (eV) | -6.543 | -5.82 | -6.81 | -5.977 | -7.40 | -6.44 |

Any further enhancement of the theoretical design of superlattices including ternary alloys, such as InAs/InAs1-xSbx SLS, requires a precise band structure modeling approach. The challenges for this task lie in the thickness of each layer and group V molar fractions (x) of all the ternary layers in the structure. While the molar fraction can take values between 0 and 1 for each ternary layer of the superlattice, the required number of tight binding fitting parameters would be too much for an empirical case-by-case fitting approach. Instead, a direct method should be proposed to generate the tight-binding parameters of bulk ternary alloys before including these alloys in the superlattice Hamiltonian.

To address that, VCA was used to model the atomic composition of the ternary alloy. This modeling consists of replacing the two substitutable group V atoms in a ternary bulk alloy in tight-binding method by a virtual atom with shared weighted average properties of the two initial group V atoms. In terms of tight-binding parameters, we first define the TBPs of ternary alloys as the weighted average of the two binary compound semiconductors, such as an extension of Vegard law. This can be written as the following:

$$V_{\mu a v c}(A_{1-x}B_xC) = V_{\mu a v c}(AC)^*(1-x) + V_{\mu a v c}(BC)^*x, \quad (18a)$$

$$E_{\mu a,c}(A_{1-x}B_xC) = E_{\mu a,c}(AC)^*(1-x) + E_{\mu a,c}(BC)^*x \quad (18b)$$

where $\mu$ and $\nu$ can refer to any orbital amongst s, p, and s*. $A$ and $B$ are the substituted ions in the III-V ternary alloy.

It is widely known that the linear Vegard's law[48] is a good approximation for the lattice constant but not for the bandgap; since experimental data shows that the bandgap of ternary alloys as a function of the composition is mostly bowed in parabolic or more complicated shape (rather than to be linear). Modeling of this non-linear behavior of the bandgap of ternary alloys is not a simple task in tight-binding method, since the exact reason for this non-monotonic bowing of the bandgap of ternary alloys is not completely understood. Some works have suggested that the non-monotonic behavior stems only from disorder effects;[57, 58] others have suggested that the phenomenological bowing may be linked to other properties of the band structure of the initial components such as the valence band offset or the lattice mismatch between components [59]. In addition, using the weighted average given in (18) for the tight-binding theory imposes a non-linear behavior of the simulated bandgap. This unwanted bowing is the result of purely model-related calculations. It can be explained by looking closely at the theoretical expression of the bandgap around the Γ-point using the tight-binding method:

$$E_c = [\frac{E_{sa} + E_{sc}}{2}] + \sqrt{[\frac{E_{sa} - E_{sc}}{2}]^2 + V_{sasc}^2}, \quad (19a)$$

$$E_v = [\frac{E_{pa} + E_{pc} + \Delta_a + \Delta_c}{2}]$$
$$+ \sqrt{[\frac{E_{pa} - E_{pc} + \Delta_a - \Delta_c}{2}]^2 + V_{xaxc}^2} \quad (19b)$$

In these expressions, the summation expressions with the terms $V_{sasc}^2$ and $V_{xaxc}^2$ inside the square roots are responsible of the non-linearity of $E_c$ and $E_v$ as a function of the group V molar fraction (x) when the parameters $E_{sa}, E_{sc}, E_{pa}, E_{pc}$ are linearly interpolated for ternary alloys using the expression (18b). Even though this non-linearity is not significant, it must be taken into the account and compensated when we pursue modeling of the bowing effect for the band gap of ternary alloys.

The first assumption we make is the shift of the bandgap in ternary alloys comes from the displacement of the lowest conduction band rather than that of the highest valence band; since the effective mass of holes in semiconductors are generally much higher than the electrons. This leads us to consider that the phenomenological bowing of a ternary alloys' bandgap could, instead be modeled by a bowing in either the s orbital $E_{sa}$ or $E_{sc}$ parameters in the tight-binding approach (depending on what type of ion is substituted in the ternary alloy being considered). This bowing is a second-order approximation, and is complementary to the previously described linear weighted average for all the binding parameters in (18):

$$E_{sv}(A_{1-x}B_xC) = E_{sv}(AC)^*(1-x) + E_{sv}(BC)^*x + b_v^* x^*(1-x) \quad (20)$$

where $v = a, c$; A and B are the substituted ions, and $b_v$ is the empirical bowing parameter. The bowing parameter $b_v$ in the virtual crystal approximation can be written as [60]:

$$b_v = k \frac{|E_{sv}^{AC} - E_{sv}^{BC}|^\lambda}{|V_{sasc}^{AC} - V_{sasc}^{BC}|} \quad (21)$$

This may not be the best way of presenting $b_v$ to fit it for ternary alloys, since it introduces two other ($k$ and $\lambda$) unknown parameters. However, the physical interpretation behind this formalism is useful. If the two binary compound semiconductors which are the basis of a ternary alloy share the same s-on-site energy, the bowing of the energy gap should be nearly equal to zero, which agrees with experimental results.

The bowing parameter $b_v$ in the VCA with linear, and second order bowing for the s-on-site energies have been fitted to the commonly agreed values of the bandgap energy at Γ-point of ternary compounds. These reference values can be found in the literature and result from previous experimental studies [61]. In this study, $\lambda = 1$ was used for all alloys.

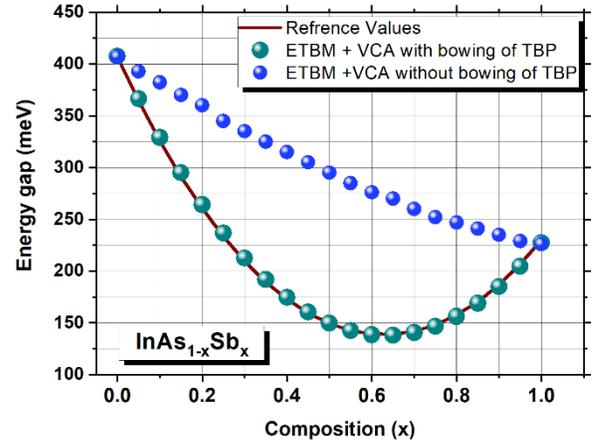

**Fig. 1.** Simulated bandgap of InAs$_{1-x}$Sb$_x$ III-V bulk ternary alloys. The brown line is the reference bandgap values.

Fig. 1 shows that the predicted bandgap energies at the Γ-point for InAs$_{1-x}$Sb$_x$ and AlAs$_{1-x}$Sb$_x$ bulk ternary alloys are in good agreement with experimental data when the virtual crystal approximation is used in combination with a second order bowing parameter for the s-on-site $E_{sv}$ energy (black dots).
In contrast, the blue dots represent the simulated bandgap values for different group V incorporation ratios (x) when a linear weighted average of the tight-binding parameters is used. This figure shows that a first order approximation already entails a bowing of the band gap as a function of the composition (x), especially for the AlAs$_{1-x}$Sb$_x$ ternary alloy as it was described in (19). This validates the choice of only bowing the s-on-site $E_s$ energy to model the bandgap of the ternary alloys using the tight-binding method.
To provide more realistic modeling for InAs/InAsSb superlattice grown by molecular beam epitaxy (MBE) in a non-ideal condition, the segregation of adjacent layers in superlattices should be considered in theoretical model for band structure calculation. Segregation occurs during a real growth by MBE and consists in the mixing of elements at heterojunctions, which slightly modifies the resulting electronic properties of the grown superlattice. Several works reported [62-67]. to describe the segregation profile of multiple group III-V elements and tried to provide understanding of physical origin of segregation. Understanding the physics behind the phenomena is crucial to select the segregation of which element to pick in our tight-binding model.

During growth by MBE, the sample's surface is alternatively sprayed by different molecular beams to form each layer of the superlattice. This molecular beam is composed of thermally activated particles that instead of being incorporated on the surface of the sample may remain inside the growth chamber following the statistical processes. These remaining particles might get incorporated in subsequent layers and introduce the segregation to the grown material. Thermal excitation of atoms allows them to break free from their formed bonds with the growing sample and become free particles in the reactor, suggesting major cause of segregation is surface desorption. To model this behavior, two-state Boltzmann distribution where a given atom is whether a free particle or tightly bonded on the sample. The Boltzmann factor linked to the probability of an atom to desorb is:

$$F(free) = \frac{1}{1 + e^{-\frac{E_{bond}}{kT}}} \quad (22)$$

This simple model can qualitatively determine which elements are more prone to segregate. Another assumption is to neglect the segregation of In and Al group III elements, since the actual MBE growth we used in this work are all group V rich and group III limited. For group V elements, the bond strength of Sb with indium is weaker than that of As. As is presented in Table III, the bond strength involving Sb, As and even Al are higher than that of bonds involving In. Taking into account the segregation of antimony only in our model can be somehow realistic considering scanning tunneling microscopy studies performed on real samples [67].

**Table III.** Bond enthalpies involving As, Sb and some group III elements (kJ/mol)

|    | Al    | Ga    | In    | Sb    | As    |
|----|-------|-------|-------|-------|-------|
| Sb | 216.3 | 192.0 | 151.9 | 299.2 | 330.5 |
| As | 202.9 | 209.6 | 201.0 | 330.5 | 382.0 |

The previous works [67] suggested a geometric progression of the segregation pattern of Sb in As-containing layers, when segregation coefficient ($R$) represents the fraction of antimony segregated, which presented as following for the segregation profile:

$$x(n) = x_i R^{n-1}(1 - R) + x_0(1 - R^n) \quad (23)$$

where $x(n)$ is the n-th arsenic layer grown after an antimony superlattice layer, $x_i$ is the initial seed of antimony and $x_0$ is the background antimony incorporation. It has been suggested[67] that the seed $x_i$ depends on the the composition of the Sb-based layer which is segregated in the subsequently-grown As-based layers (such as InAs). The segregation factor $R$ on the other hand depends on the physical properties of the grown layers, as an Sb atom will more likely be incorporated in a layer where the bonding energies are more prone to include an Sb atom. This can result in raising the value of R for that layer. As our model is designed to be able to simulate the bandgap of many different superlattices, including a wide range of Arsenic-based alloys such as InAs and AlAs, we chose to consider a single choice of average values $x_i = 0.4$, $R = 0.67$ and $x_0 = 0.07$ for all the segregated layers, as the shift of the simulated bandgap resulting of segregation is far less important than the choice of TBP or valence band offset (VBO).

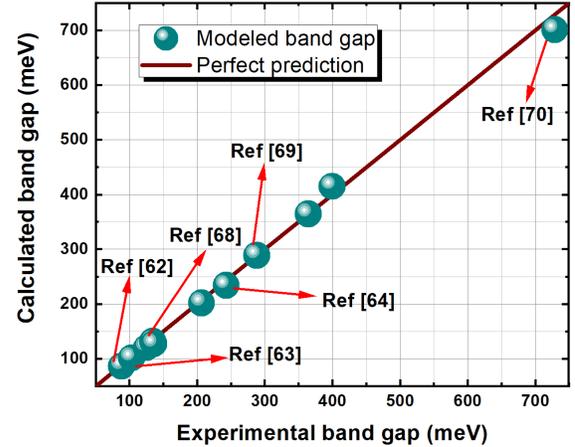

**Fig. 2.** Comparison between the simulated bandgap of different superlattice designs of ternary alloys and measured bandgap energy of corresponding experimental superlattices grown with MBE.

The choice of TBPs shows good agreement with experimental results (Fig. 1). Additionally, the model has been employed successfully to predict the band structure of several different superlattice designs including III-V ternary alloys. Fig. 2 shows good agreement between the predicted bandgap energy using our ETBM model presented in this work and the experimental values for several different superlattices with cut-off wavelengths ranging from 1.5 to >14 μm at 77 K.

All the data for experimental data was taken from MBE grown samples across the different range of the wavelength. We assumed that the segregation pattern is the same for all samples. For the InAs/InAs$_{1-x}$Sb$_x$ type-II superlattices (Fig 3), the electrons and holes are spatially separated in InAs and InAs$_{1-x}$Sb$_x$ quantum wells and the effective bandgap of the superlattice can be engineered by changing the layer thicknesses and antimony molar fraction of the InAs$_{1-x}$Sb$_x$ layer. It has been demonstrated that the bandgap of this type of superlattices can be widely tuned from a few tens to a few hundreds of meV.

Fig. 3 presents the simulation results for [(InAs)$_{35}$ – (InAs$_{0.57}$Sb$_{0.43}$)$_{12}$] superlattice design which is in a good agreement with previous experimental work [25] for identical design. Other comparison with experimental results confirming the same outcome. The modelling calculation shows 83 meV for the bandgap, which is corresponding to 14.93 μm at 77 K, which is very close to 50% cut-off frequency reported in the experimental result [16].

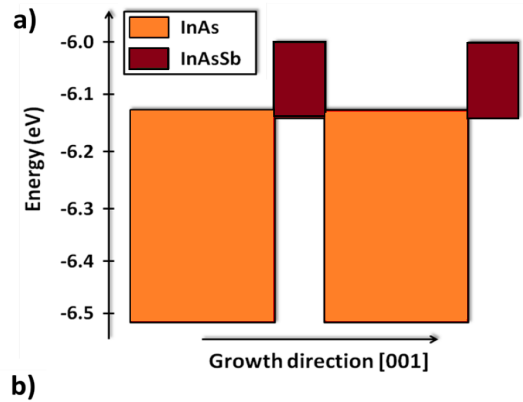

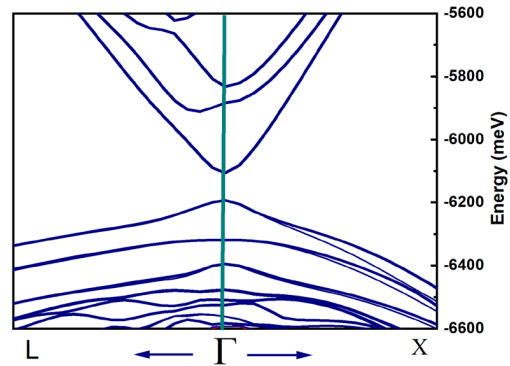

**Fig. 3.** (a) $[(InAs)_{35} - (InAs_{0.57}Sb_{0.43})_{12}]$ superlattice design and (b) simulated band structure around the Γ-point of the superlattice using a $sp^3s^*$ tight-binding model. The colored regions represent the energy gap of each layer in the structure.

In this work the ETBM approach as a quantum mechanical based modelling, was used to develop modeling on bandgap calculation for $InAs/InAs_{1-x}Sb_x$ SLS material. In this approach, VCA was implemented to calculate the precise ETBM fitting parameters of ternary compounds along with considering of the antimony segregation factor during the superlattice growths. The ETBM modeling results were compared with experimental results for different variations in designs of bulk ternary and superlattice design with acceptable agreement, suggesting suitability of the approach for more complex designs for future works.